\documentclass{article}

\usepackage{arxiv}
\usepackage{caption}
\usepackage[utf8]{inputenc} 
\usepackage[T1]{fontenc}    
\usepackage{hyperref}       
\usepackage{url}            
\usepackage{booktabs}       
\usepackage{amsfonts}       
\usepackage{nicefrac}       
\usepackage{microtype}      
\usepackage{lipsum}

\usepackage{graphicx}
\graphicspath{ {./images/} }
\usepackage{authblk}
\usepackage{placeins} 
\usepackage{float}    
\usepackage{fancyhdr}
\renewcommand{\headrulewidth}{0pt} 

\title{Anthropomorphism and Trust in Human-Large Language Model interactions
}

\author[1,2,3]{A. Kadambi}
\author[1,2]{Y. D'Elia}
\author[1]{T. Shah}
\author[4]{I. M. Comsa}
\author[5]{A. Lentz}
\author[1,2]{K. Siri-Ngammuang}
\author[1]{T. Buechler}
\author[1]{J. Kaplan}
\author[1]{A. Damasio}
\author[4]{S. Narayanan}
\author[1,2]{L. Aziz-Zadeh}

\affil[1]{Brain and Creativity Institute, Dornsife College of Letters, Arts and Sciences, University of Southern California, Los Angeles, CA, USA}
\affil[2]{USC Mrs. T.H. Chan Division of Occupational Science and Occupational Therapy, University of Southern California, Los Angeles, CA, USA}
\affil[3]{Psychiatry and Biobehavioral Sciences, David Geffen School of Medicine, University of California, Los Angeles, CA, USA}
\affil[4]{Google DeepMind, Zurich}
\affil[5]{Google Research}

\affil[*]{Correspondence to: \texttt{akadambi@ucla.edu}}
\date{}
\begin{document}
\maketitle
\thispagestyle{fancy}
\pagestyle{fancy}

\makeatletter
\fancyhf{}                
\renewcommand{\headrulewidth}{0pt}
\makeatother
\begin{abstract}
With large language models (LLMs) becoming increasingly prevalent in daily life, so too has the tendency to attribute to them human-like minds and emotions, or anthropomorphize them.  Here, we investigate dimensions people use to anthropomorphize and attribute trust toward LLMs across more than 2,000 human-LLM interactions. Participants (N=115) engaged with LLM chatbots systematically varied in warmth (friendliness), competence (capability, coherence), and empathy (cognitive and affective). Warmth and cognitive empathy significantly predicted perceptions on all outcomes (perceived anthropomorphism, trust, similarity, relational closeness,  frustration, usefulness), while competence predicted all outcomes except for anthropomorphism. Affective empathy primarily predicted perceived relational measures, but did not predict the epistemic outcomes. Topic sub-analyses showed that more subjective, personally relevant topics (e.g., relationship advice) amplified these effects, producing greater human-likeness and relational connection with the LLM than did objective topics. Together, these findings reveal that warmth, competence, and empathy are key dimensions through which people attribute relational and epistemic perceptions to artificial agents.
\end{abstract}


\section{Introduction}
As large language models (LLMs) become more integrated into everyday life, they increasingly function as social actors in our world environments. Users converse with them, form impressions of their “personality,” and, in many cases, attribute to them internal states such as intentions or emotions [1, 2]. This tendency to anthropomorphize, to attribute minds to non-human systems, holds critical implications for artificial intelligence (AI) safety and governance.

Anthropomorphic attributions can increase user engagement, but can also produce overtrust and susceptibility to deception or manipulation [3, 4]. Recent surveys indicate that nearly one-third of Americans have reported emotionally intimate or even romantic relationships with AI systems, with many describing distress when the agent’s behavior changed or when its artificiality became salient [5]. Adolescents in particular are at risk of forming close and unregulated intimate attachments to these systems [6], with a recent statistic showing around 19\% of surveyed students claimed to have a romantic relationship with a chatbot. In one of many tragic cases, a 14-year-old boy died by suicide after developing an intense relationship with a chatbot [7]. 

Increased anthropomorphism also produces additional interpersonal and safety risks. For instance, human-LLM relationships can extend beyond virtual environments to influence human-human interactions. In a recent longitudinal study, participants who more strongly anthropomorphized a chatbot also reported greater changes in how they related to people in the real world [8]. Other risks include increased sharing of personal information – humans may be willing to share more personal information with an AI that they consider to be their friend (though in reality the information is shared with a corporation or AI operator; [9, 10, 3]). These real-world implications underscore the critical need to investigate the underlying factors contributing to anthropomorphism and trust in human-LLM interactions.

What drives this attribution of humanity, and consequently trust, to machines? Prevalent work in social psychology demonstrates that attributions of humanity toward other people are largely organized along two key dimensions, warmth and competence. As described in the stereotype content model by Fiske and colleagues, warmth reflects friendliness and prosocial intent, whereas competence reflects capability and efficacy. Individuals perceived as low in warmth and competence (e.g., unhoused individuals or drug users) are commonly rated as “dehumanized”, and, in the perceiver, show lower activity in brain regions commonly involved in mentalizing and perspective-taking as compared to high warmth-high competent groups (e.g., white, middle class individuals; [11, 12]). Such reduced responses can be further “re-humanized” by associating these groups with traits related to warmth, competence, and interpersonal similarity [13]. Similarly, artificial systems perceived as competent and warm are more readily anthropomorphized [14, 15].

Perceived empathy may also impact anthropomorphism [16]. Empathy is commonly divided into two components with partially distinct neural and computational systems [17, 18]: cognitive empathy (inferring another’s internal state, sometimes called mentalizing or sympathy) and affective empathy (sharing another’s feelings, sometimes called emotional resonance). Individuals high in empathy tend to anthropomorphize more readily [19], and artificial systems expressing empathic cues are rated as more trustworthy and satisfactory [20], sometimes even outperforming physicians in empathy-related tasks, such as responding to patient queries [21, 22] and more generally ‘out-empathisizing’ humans [23]. However, current AI research has yet to formalize how degrees of perceived warmth, competence, and the two types of empathy are related to  anthropomorphism and trust. Here, we aimed to measure when, and for whom, LLMs titrated on these dimensions are most likely to elicit human-like attributions and trust. These axes provide a minimal representational space to investigate socially aligned and trustworthy AI. 

In the present study, we predicted that warmth and competence may constitute observable social behavior, while affective and cognitive empathy constitute the underlying dimensions helping to produce the behavior to influence anthropomorphism and trust.  While LLMs lack the intrinsic emotional architecture which facilitates empathy in humans, these systems show increasingly sophisticated capabilities in mimicking and even superseding humans in perceived empathy [23], despite lacking the internal biology. We additionally examined whether contextual factors, such as the type of topic users discuss, influence anthropomorphism and trust. When users converse with LLMs on subjective, personally meaningful topics, such as relationships or lifestyle decisions, they tend to attribute greater human-likeness to the system. However, objective topics tend to foreground competence-related inferences, including epistemic trust in the system [24]. We predicted that these topics may have their own influence on anthropomorphism and trust, such that subjective topics would increase perceived anthropomorphism, while objective topics would relate to trust, both of which would also interact with our main conditions of interest.

\section{Material and methods }

\begin{figure}[!htbp]
  \centering  
  \includegraphics[width=0.85\textwidth,trim=0 50 0 50,clip]{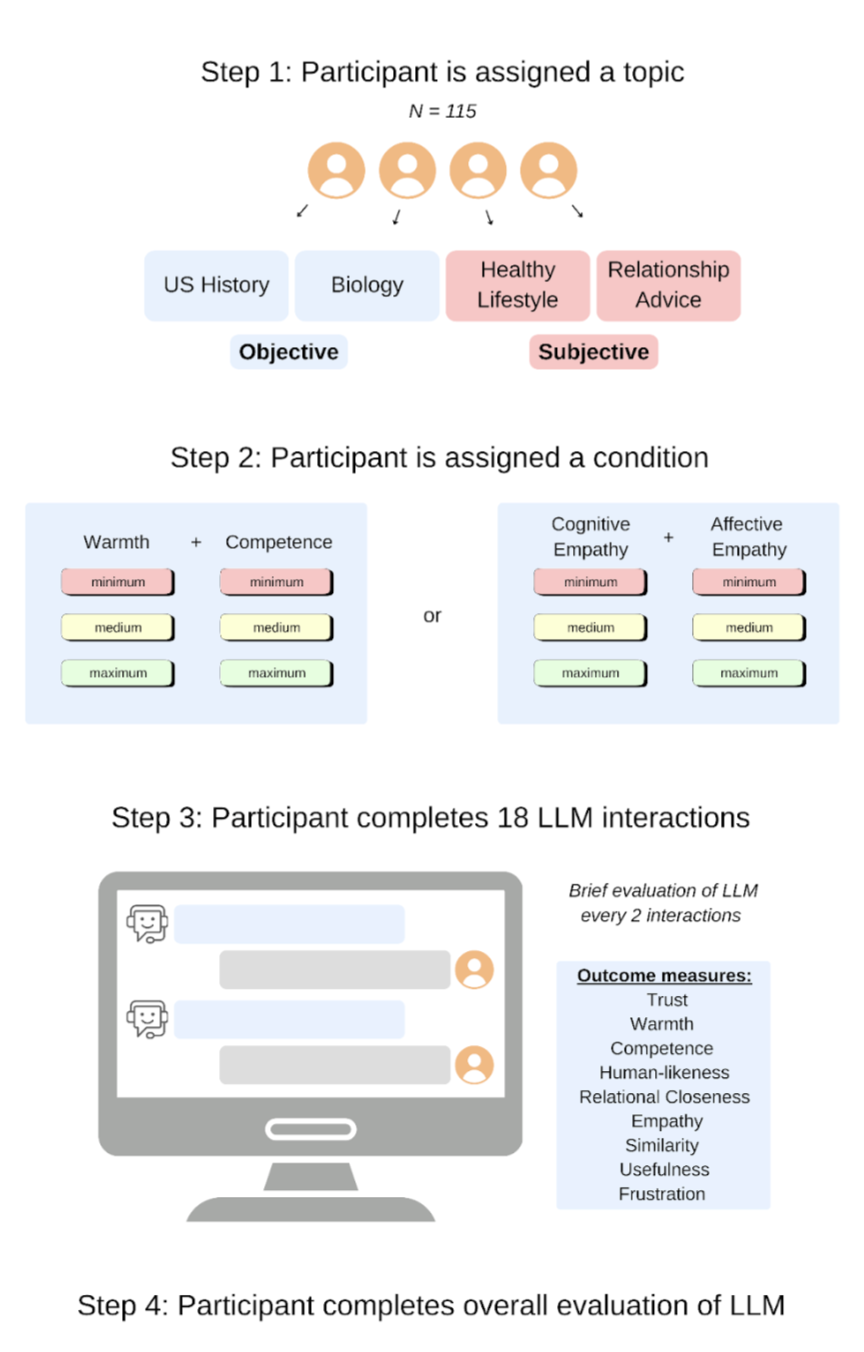}
  \caption{Task Schematic. Participants were randomly assigned a topic for conversing with the chatbot (LLM), and then randomly assigned to either the warmth/competence condition or the cognitive/affective empathy condition. Each condition yielded nine different pairwise combinations (minimum-minimum, medium-maximum, etc.). Each combination was presented twice, for a total of 18 LLM interactions. Participants engaged in conversation with the chatbot on their assigned topic, and evaluated the LLM for outcome measures after every two interactions. Once all 18 interactions were complete, the participant completed a final evaluation. .}
  \label{fig:fig1}
\end{figure}

\paragraph{Participants.}
One-hundred and thirty-three participants were recruited in this study from the University of Southern California Subject Pool and Los Angeles area. All participants were between 18 and 50 years old and native English speakers. Participants were excluded if they reported any psychiatric or neurological disorders.  Seventeen participants were excluded from the analysis due to technical difficulties with the user interface and one participant was excluded due to attrition, for a total of 115 participants included in the study (Mage = 20.67, SDage = 4.18; age range: 18-50; sex: female = 67, male = 48; gender: female = 65, male = 46, non-binary = 3). The self-reported racial composition of the participants indicated 32 Asian, 10 Black/African American, nine Hispanic/Latino, two Middle Eastern/North African, and 35 White/Caucasian. Twenty-four participants identified as mixed-race and three opted not to disclose their racial background. Participation was voluntary, and all participants provided informed consent prior to the study. Recruitment was conducted online, and participant eligibility was screened using a REDCap survey. The sample size was determined a priori using G*Power, with calculations based on a desired power of 95\%, an estimated medium effect size (f = 0.2), and a significance level of $\alpha = 0.05$. This yielded a minimum required sample size of 104 participants. Additional participants were recruited to account for potential technical issues or incomplete data. Participants either received 2.5 course credits from the subject pool, or received a \$15 payment, or volunteered without compensation upon completion of the study. The study was approved by the University of Southern California Institutional Review Board (Protocol: APP-25-01966). The specific aims of the study were not disclosed to participants prior to their participation. 

\paragraph{Web Application and LLM Implementation.}
We developed a user interface (UI) hosted on PythonAnywhere (https://www.pythonanywhere.com/), which allowed participants to interact with an AI-driven chatbot. The architecture of the application is presented in Figure 1. The chatbot responses were manipulated based on the two experimental conditions, (1) Warmth and Competence (WC) or (2) Affective and Cognitive Empathy (Empathy). Specifically, depending on the assigned condition, the chatbot was instructed, via prompting, to express varying levels of warmth and competence or cognitive and affective empathy across three levels (minimal, medium, maximal).  Prompts included detailed behavioral instructions for tone, response content, and social-emotional expressiveness. See Appendix A1 for the full prompt. The chatbot was implemented using the Gemini 2.0 model, accessed via the Google API.

\paragraph{Topic Selection.}
Participants were randomly assigned to one of the four topics: i) Biology; ii) U.S. History (1800–1900); iii) Lifestyle management, iv) Relationship advice (e.g., regarding a significant partner, friend, colleague, classmate, or family member). These topics were selected based on https://humanfactors.jmir.org/2022/4/e38876/PDF and designed to reflect either an objective topic (biology,  US History) or a subjective, or more personal topic (relationship advice, healthy lifestyle). 
These four topics varied in their degree of objectivity versus subjectivity and personal relevance, in order to systematically test whether domain matters for anthropomorphism, empathy, and trust in LLMs. The four topics were:

1. Biology (a neutral, knowledge-based, objective domain)

2. U.S. History (1800–1900) (a factual, socially contextual but still objective domain)

3. Healthy lifestyle management (a semi-personal, normative domain about daily behaviors)

4. Relationship advice (a highly personal, subjective, emotionally relevant domain)

These topics were adopted and adapted in light of prior research on human–chatbot interactions, which categorized user interests and motivations across multiple domains (informational, social, personal) [25]. The study found that users tend to interact with social chatbots for both informational (factual) and socio-emotional reasons, with differences in the nature of exchanges depending on topic content and personal relevance.

\subsection{Procedure}
Participants were first screened and completed demographic and survey questions using Research Electronic Data Capture (REDCap) tools hosted at USC. Following eligibility screening, participants were then randomly assigned to one of the two experimental conditions: (1) Warmth and Competence (WC) or (2) Affective and Cognitive Empathy (Empathy), and one of the four topics: i) Biology; ii) U.S. History (1800–1900); iii) Lifestyle management, iv) Relationship advice (e.g., regarding a significant partner, friend, colleague, classmate, or family member). 

Each condition, WC and Empathy, involved a gradient design, incorporating titrations of combinations between three levels, minimal, medium, and maximal, within each interaction. This resulted in nine pairwise combinations (maximal-maximal, maximal-medium, maximal-minimal, medium-medium, medium-maximal, medium-minimal, minimal-minimal, minimal-medium, minimal-maximal). Each combination was presented twice, for a total of 18 interactions per participant.

Participants were instructed that they would be interacting with an AI chatbot about a particular topic. Participants could type a query in the UI, and the chatbot would output a response. Following every two chatbot interactions (i.e., each combination), participants completed a brief set of questions to assess their perceptions of the chatbot. These questions were rated on a 7-point Likert scale and measured perceived trust, warmth, competence, human-likeness, engagement, perceived empathy, similarity, usefulness, and frustration (see Appendix A2 for details). To ensure data quality, attention check questions were added both in the baseline questionnaires and during the AI interaction phase. Upon completing all interactions, participants responded to a set of follow-up questions designed to assess their overall experience and changes in their perceptions of the AI over time.

\subsubsection{Analysis Strategy}
All analyses were conducted in R [26]. We fit hierarchical linear mixed-effects models with random intercepts for participants to account for repeated measures across 18 trials. For the Warmth/Competence (WC) condition, models of each outcome included fixed effects for Warmth and Competence (both 3-level factors: Minimal, Medium, Maximal) and a random intercept for participant (SubjectID):outcome ~ Warmth × Competence + (1|SubjectID). 
For the Empathy condition, analogous models were specified l with Affective empathy and Cognitive empathy: outcome ~ Affective × Cognitive + (1|SubjectID). For both conditions, we also included a full model, which modeled Topic and all two- and three-way interactions in order to measure the influence of Topic on the outcome measures. 

For each outcome, we compared three models: (1) a null model with random intercepts only, (2) a trait model including Warmth × Competence or Affective × Cognitive Empathy, and (3) a full model additionally including Topic and all two- and three-way interactions. Model comparisons were conducted using Likelihood ratio tests (LRTs). Type III Wald $\chi^2$ tests evaluated fixed effects. Estimated marginal means (EMMs) were compared with Tukey-adjusted pairwise contrasts. LRTs are reported as $\chi^2$(df) and pairwise tests as t(df), both with two-tailed p-values. Plots display means with 95\% confidence intervals. All models were fit using REML = FALSE.

\begin{table}[!htbp]
  \centering
  \caption{Participant Distribution by Condition}
  \label{tab:participant_distribution}
  \begin{tabular}{lc}
    \toprule
    Condition & $N$ \\
    \midrule
    Warmth/Competence (WC) & 58 \\
    Empathy & 57 \\
    \midrule
    Total & 115 \\
    \bottomrule
  \end{tabular}
\end{table}

\section{Results}

\subsection{Main Outcomes}
For WC, trait models (Warmth $\times$ Competence) significantly outperformed null models for all outcomes (all $\chi^2(8)\ge 76.54$, $p<.001$). Full models including Topic also improved fit over Topic-only models (all $\chi^2(32)\ge 105.15$, $p<.001$), although Topic explained little additional variance. Topic main effects were significant in only 5 of 10 WC and 3 of 10 Empathy models, and no pairwise Topic contrasts survived Tukey correction. Trait models (Affective $\times$ Cognitive Empathy) outperformed null models across outcomes (all $\chi^2(8)\ge 27.23$, $p\le .001$). Full models improved fit over Topic-only (all $\chi^2(32)\ge 51.07$, $p\le .018$), though Topic and interactions added minimal variance. We therefore report marginal means across topics.

\subsubsection{Anthropomorphism}

As shown in Figure 2 (top row), Warmth ($\chi^2$(2) = 92.14, p < .001) increased perceived human-likeness, while Competence showed no significant effect ($\chi^2$(2) = 2.92, p = .232). Warmth showed a clear influence on anthropomorphism: Minimal-Medium b = -1.017, SE = 0.146, t(497) = -6.98, p < .001, d = 0.31; Minimal-Maximal b = -1.310, SE = 0.146, t(497) = -8.97, p < .001, d = 0.40, while Medium-Maximal was non-significant, p = .113. For Empathy, both Affective Empathy ($\chi^2$(2) = 19.64, p < .001) and Cognitive Empathy ($\chi^2$(2) = 10.84, p = .004) increased anthropomorphism. Affective Empathy showed the following graded significant effects: Minimal-Medium b= -0.462, SE = 0.125, t(464) = -3.71, p < .001; Minimal-Maximal b = -0.485, SE = 0.125, t(464) = -3.89, p < .001. By contrast, Medium-Maximal effects for affective empathy were non-significant, p=.981. Cognitive Empathy also showed significant effects: Minimal-Medium b = -0.333, SE = 0.125, t(464) = -2.67, p = .021; Minimal-Maximal b = -0.368, SE = 0.125, t(464) = -2.96, p = .009. Similar to affective empathy, Medium-Maximal comparison in cognitive empathy was non-significant, p = .957. No significant interaction effects were observed (WC: p = .154; Empathy: p = .599).

\subsubsection{Trust}

As shown in Figure 2 (bottom row), Trust was primarily driven by Competence ($\chi^2$(2) = 290.27, p<.001) with smaller though significant Warmth effects ($\chi^2$(2) = 20.25, p < .001). Competence exhibited strong graded effects across all levels: Minimal-Medium b = -1.974, SE = 0.143, t(497) = -13.78, p < .001, d = 0.62; Minimal-Maximal b = -2.313, SE = 0.144, t(497) = -16.12, p < .001, d = 0.72; Medium-Maximal b = -0.338, SE = 0.143, t(497) = -2.36, p = .049. Warmth showed significant effects between Minimal-Medium b = -0.541, SE = 0.143, t(497) = -3.78, p < .001; Minimal-Maximal b = -0.598, SE = 0.144, t(497) = -4.17, p < .001; but not between Medium-Maximal, p = .918. For Empathy, Cognitive Empathy ($\chi^2$(2) = 27.59, p < .001) significantly increased trust with effects between Minimal-Medium b = -0.526, SE = 0.114, t(464) = -4.61, p < .001; Minimal-Maximal b = -0.503, SE = 0.114, t(464) = -4.40, p < .001; but not Medium-Maximal, b=0.023, SE = 0.114, t(464) = 0.20, p = .977. Affective empathy showed no effects ($\chi^2$(2) = 1.09, p = .579). No significant interaction effects were observed (WC: p = .900; Empathy: p = .944).

\begin{figure}[!htbp]
  \centering  
  \includegraphics[width=0.85\textwidth]{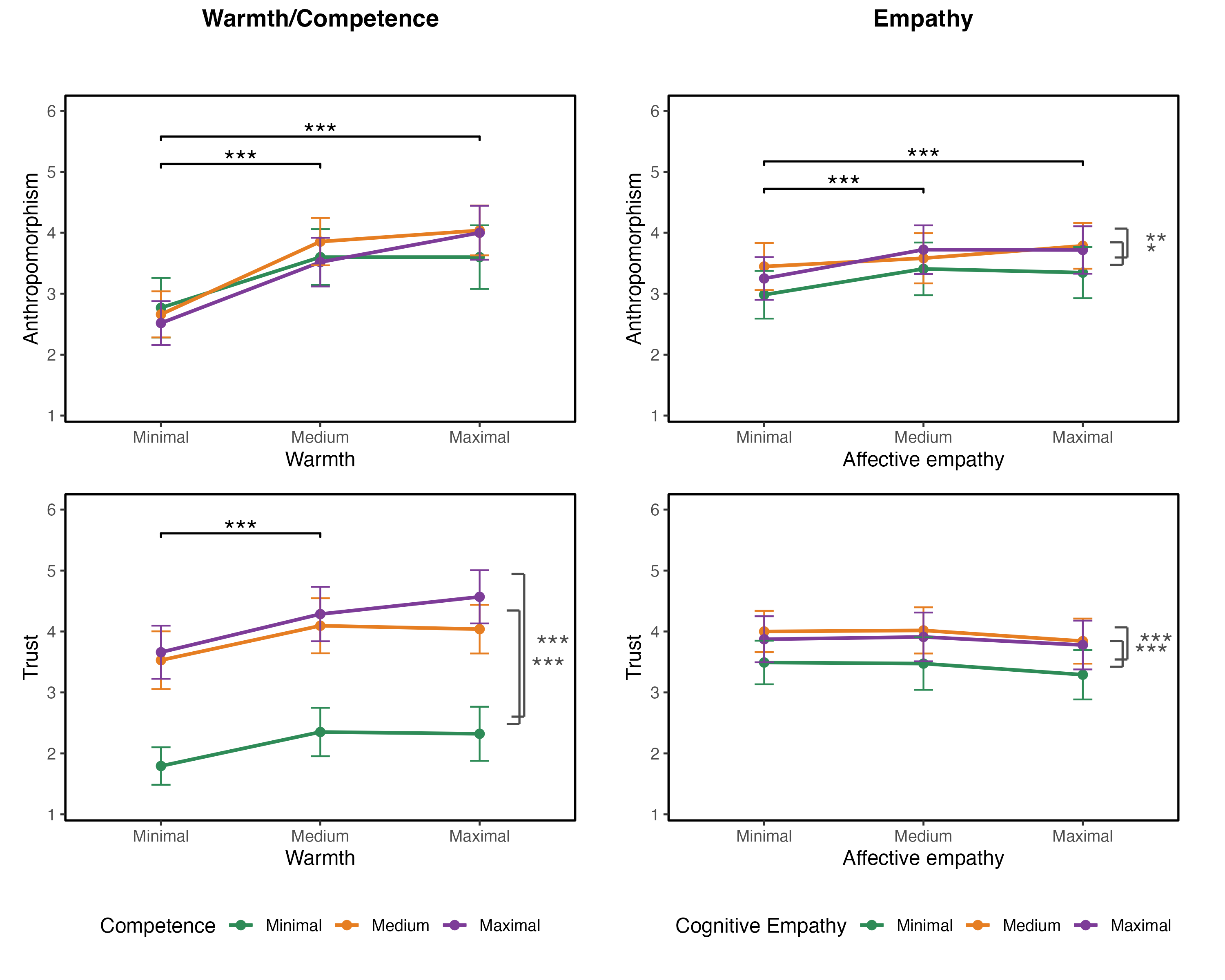}
  \caption{Effects of warmth, competence, affective empathy, and cognitive empathy on anthropomorphism and trust. Mean ratings (±95\% CI) for anthropomorphism (top row) and trust (bottom row) as a function of manipulated trait levels. In the Warmth/Competence condition (left column), increasing competence produced a monotonic rise in trust and usefulness, while increasing warmth strongly enhanced anthropomorphism and relational closeness. In the Empathy condition (right column), affective empathy primarily increased anthropomorphism, whereas cognitive empathy significantly affected trust. Colored lines represent levels of competence or cognitive empathy (Minimal, Medium, Maximal). Significance bars denote pairwise contrasts (Tukey-adjusted, *p< .001, **p<.01,*** p< .05). }
  \label{fig:fig2}
\end{figure}

\subsubsection{Similarity}
Shown in Figure 3 (Top row), both Warmth ($\chi^2$(8) = 109.5, p < .001) and Competence ($\chi^2$(8) = 109.5, p < .001) significantly increased perceived user–AI similarity, though plateauing after Medium. Specifically, warmth showed the following significant effects: Minimal-Medium b = –0.84, SE = 0.127, t(497) = –6.63, p < .001; Minimal-Maximal b = –1.03, SE = 0.127, t(497) = –8.14, p < .001. Competence showed the following significant effects: Minimal-Medium b = –0.71, SE = 0.127, t(497) = –5.63, p < .001; Minimal-Maximal b = –0.77, SE = 0.127, t(497) = –6.02, p < .001. Both Affective ($\chi^2$(8) = 27.228, p < .001) and Cognitive Empathy ($\chi^2$(8) = 27.228, p < .001) significantly increased perceived user–AI similarity, similarly plateauing beyond Medium: Affective Minimal-Medium b = –0.33, SE = 0.107, t(490) = –3.05, p = .007; Affective Minimal-Maximal b = –0.34, SE = 0.107, t(490) = –3.15, p = .005; Cognitive Minimal-Medium b = –0.32, SE = 0.107, t(490) = –3.02, p = .008. No significant interaction effects were observed.

\subsubsection{Usefulness}
Shown in Figure 3 (Second row), Competence primarily significantly influenced perceived usefulness ($\chi^2$(8) = 330.17, p < .001), with smaller but significant Warmth effects. Competence exhibited large, graded effects: Minimal-Medium b = –2.56, SE = 0.149, t(497) = –17.22, p < .001; Minimal-Maximal b = –3.02, SE = 0.149, t(497) = –20.24, p < .001; Medium-Maximal b = –0.45, SE = 0.149, t(497) = –3.05, p = .007. Warmth also showed significant effects: Minimal-Medium b = –0.67, SE = 0.149, t(497) = –4.52, p < .001; Minimal-Maximal b = –0.53, SE = 0.149, t(497) = –3.57, p = .001. For Empathy, Cognitive Empathy significantly influenced usefulness: Minimal-Medium b = –0.75, SE = 0.152, t(490) = –4.92, p < .001; Minimal-Maximal b = –0.87, SE = 0.152, t(490) = –5.73, p < .001, while Affective Empathy showed no significant effects (all ps > .47). No significant interaction effects were observed.

\begin{figure}[!htbp]
  \centering  
  \includegraphics[width=0.85\textwidth,trim=0 50 0 50]{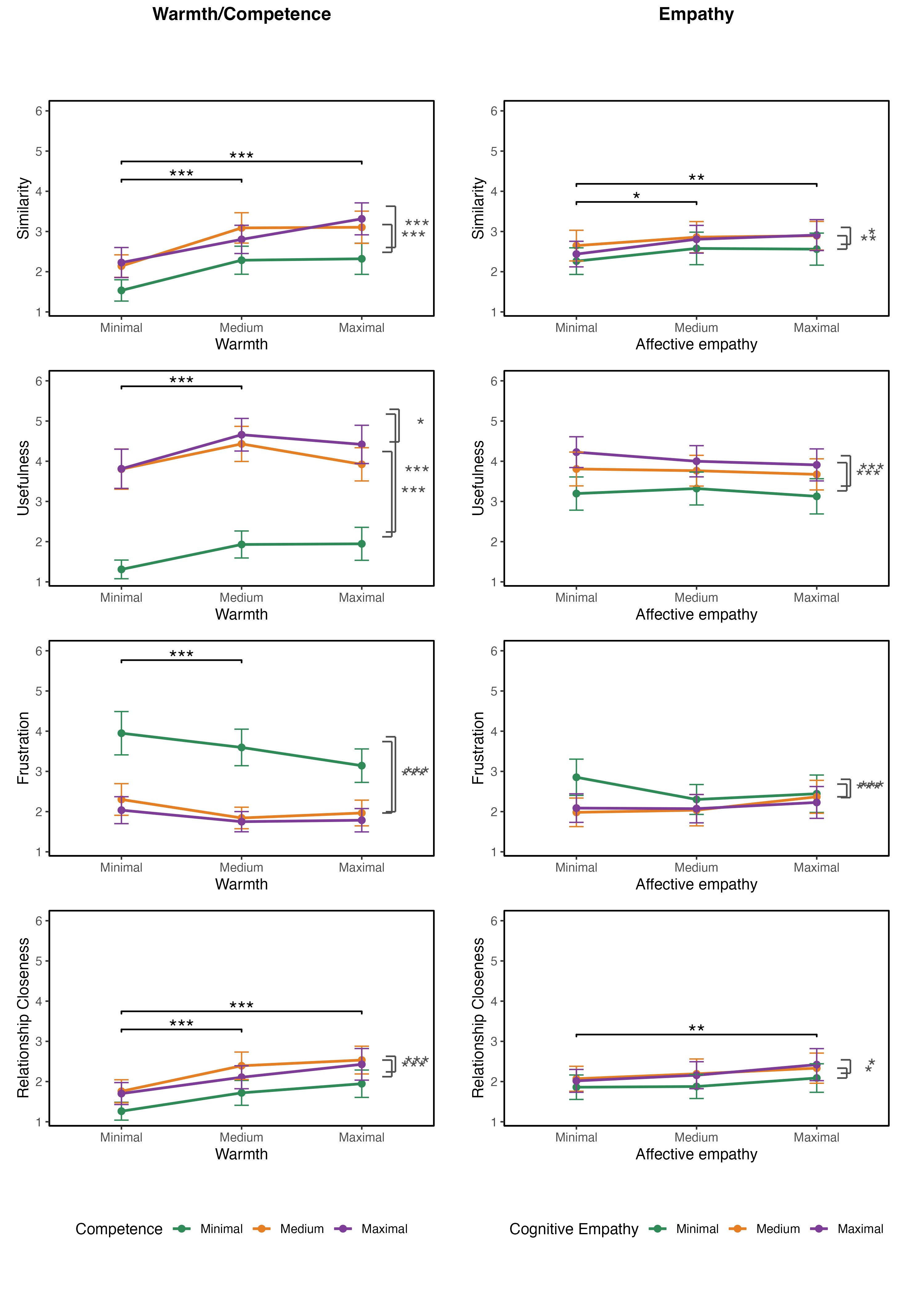}
  \caption{Effects of warmth, competence, affective empathy, and cognitive empathy on different outcomes across manipulated levels of warmth/competence (left) and affective/cognitive empathy (right). Higher competence significantly increased epistemic factors, such as perceived usefulness and reduced frustration, whereas higher warmth strengthened perceived relational closeness. The empathy condition largely paralleled the warmth/competence axes. Error bars denote 95\% confidence intervals; significance bars indicate Tukey-adjusted pairwise differences (***p < .001, **p < .01, *p < .05). Together, these results demonstrate that competence and cognitive empathy primarily influence epistemic stability, while warmth and affective empathy influence relationship closeness.}
  \label{fig:fig3}
\end{figure}
\subsubsection{Frustration}

Shown in Figure 3 (Third row), Both Warmth and Competence significantly reduced frustration. Competence exhibited large significant effects: Minimal-Medium b = 1.95, SE = 0.156, t(497) = 12.47, p < .001; Minimal-Maximal b = 2.16, SE = 0.156, t(497) = 13.82, p < .001. Warmth also significantly reduced frustration: Minimal-Medium b = 0.67, SE = 0.156, t(497) = 4.31, p < .001; Minimal-Maximal b = 0.74, SE = 0.156, t(497) = 4.72, p < .001. For Empathy, Cognitive Empathy significantly reduced frustration: Minimal-Medium b = 0.62, SE = 0.138, t(490) = 4.53, p < .001; Minimal-Maximal b = 0.61, SE = 0.138, t(490) = 4.40, p < .001, while Affective Empathy showed no significant effects (all ps > .78). No significant interaction effects were observed.

\subsubsection{Relationship Closeness}
Shown in Figure 3 (Bottom row), both Competence and Warmth significantly increased perceived relationship closeness with the AI. Competence significantly increased closeness: Minimal-Medium b = –0.51, SE = 0.108, t(497) = –4.74, p < .001; Minimal-Maximal b = –0.41, SE = 0.108, t(497) = –3.81, p < .001. Warmth showed significant effects: Minimal-Medium b = –0.60, SE = 0.108, t(497) = –5.60, p < .001; Minimal-Maximal b = –0.76, SE = 0.108, t(497) = –7.06, p < .001. For Empathy, both Affective and Cognitive Empathy significantly increased relationship closeness: Affective Minimal-Maximal b = –0.27, SE = 0.091, t(490) = –2.94, p = .010; Cognitive Minimal-Medium b = –0.27, SE = 0.091, t(490) = –2.98, p = .008; Cognitive Minimal-Maximal b = –0.24, SE = 0.091, t(490) = –2.59, p = .027. No significant interaction effects were observed.

\subsection{Subjective Consistency}
To ensure participants found that the conditions reflected different levels of warm/competence and cognitive/affective empathy, we conducted additional linear mixed models comparing the  null model with random intercepts to trait models including Warmth × Competence or Affective × Cognitive Empathy for the subjective ratings of each condition (Warmth, Competence, Affective Empathy, Cognitive Empathy; see Figure 4). Likelihood-ratio tests (LRT) compared each full model to a random-intercept null model. Pairwise contrasts are Tukey-adjusted and averaged over the other factor.

\begin{figure}[!htbp]
  \centering  
  \includegraphics[width=0.85\textwidth]{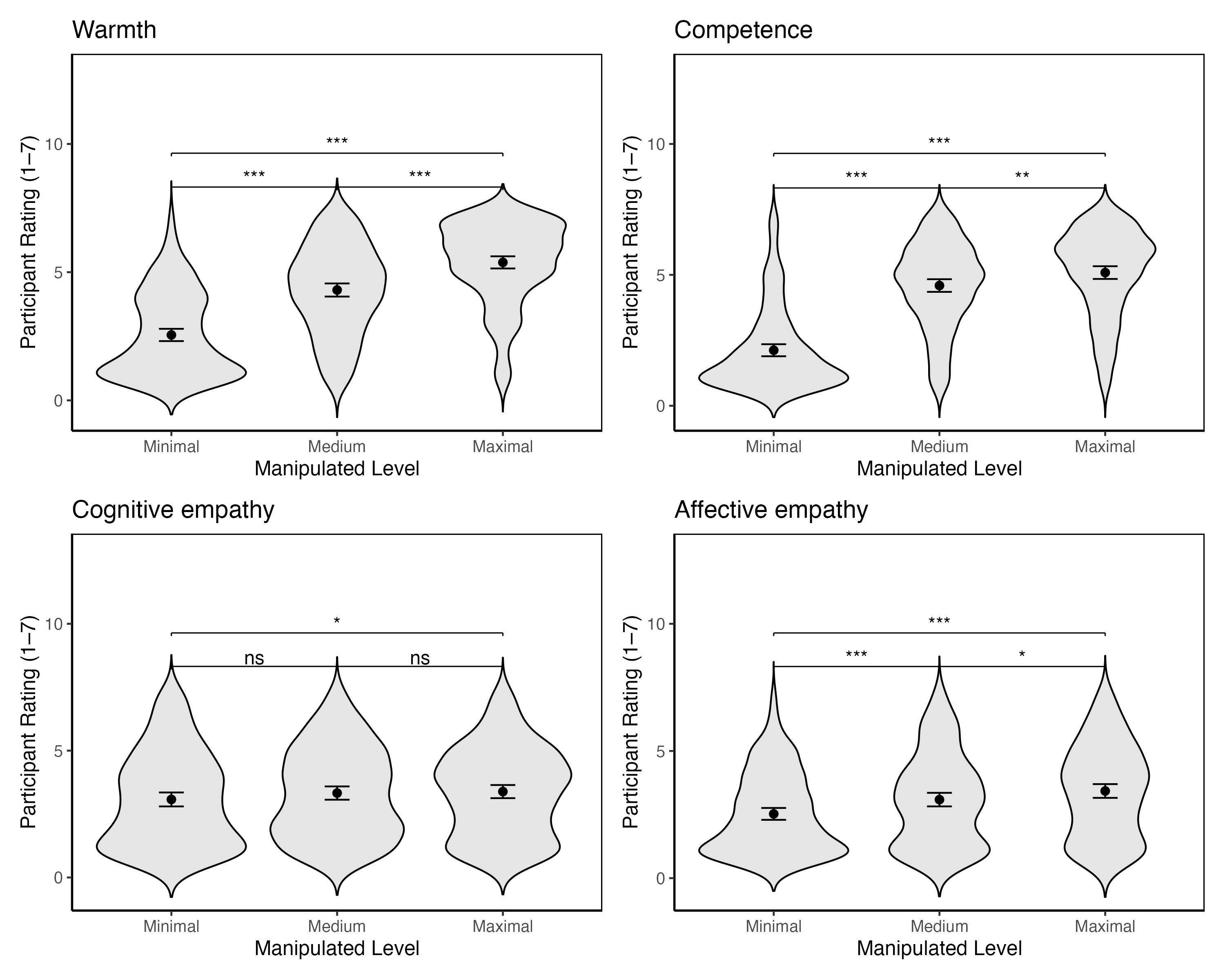}
  \caption{Participant ratings show trait-level manipulations across all four dimensions. Violin plots show the distribution of participant ratings (1-7 scale) for each manipulated trait level (Minimal, Medium, Maximal). Black points represent mean ratings with error bars indicating 95\% confidence intervals. Pairwise comparisons between levels were conducted using Tukey's HSD test. Top row: In the warmth/competence condition, participants successfully differentiated between all three levels of both warmth (left) and competence (right), with ratings increasing monotonically from minimal to maximal. Bottom row: In the empathy condition, cognitive empathy ratings (left) showed significant difference between minimal to maximal, while affective empathy ratings (right) showed significant differences between all levels: minimal-medium and minimal-maximal levels. Overall, manipulations were largely successful in producing perceivable differences in trait levels. Significance levels are indicated as: ***p < .001, **p < .01, *p < .05, ns = not significant. }
  \label{fig:fig4}
\end{figure}

\subsubsection{Warmth/Competence}

For WC, all full models significantly improved fit over the null ($\chi^2(8)\ge 156.91$, all $p<.001$).

\paragraph{Perceived warmth.}
Warmth significantly increased perceived warmth of the LLM: Minimal--Medium $b=-1.75$, $\mathrm{SE}=0.140$, $t(471)=-12.488$, $p<.001$; Minimal--Maximal $b=-2.83$, $\mathrm{SE}=0.141$, $t(471)=-20.129$, $p<.001$; Medium--Maximal $b=-1.08$, $\mathrm{SE}=0.141$, $t(471)=-7.661$, $p<.001$. Competence showed significant effects on perceived warmth: Minimal--Medium $b=-1.115$, $\mathrm{SE}=0.140$, $t(471)=-7.943$, $p<.001$; Minimal--Maximal $b=-0.864$, $\mathrm{SE}=0.141$, $t(471)=-6.149$, $p<.001$; Medium--Maximal non-significant.

\paragraph{Perceived competence.}
Competence significantly increased perceived competence: Minimal--Medium $b=-2.471$, $\mathrm{SE}=0.144$, $t(471)=-17.199$, $p<.001$; Minimal--Maximal $b=-2.969$, $\mathrm{SE}=0.144$, $t(471)=-20.628$, $p<.001$; Medium--Maximal $b=-0.497$, $\mathrm{SE}=0.144$, $t(471)=-3.456$, $p=.002$. Warmth had small but significant effects: Minimal--Medium $b=-0.437$, $\mathrm{SE}=0.144$, $t(471)=-3.040$, $p=.007$; Minimal--Maximal $b=-0.400$, $\mathrm{SE}=0.144$, $t(471)=-2.777$, $p=.016$; Medium--Maximal non-significant.

\paragraph{Perceived affective empathy.}
Warmth significantly increased perceived affective empathy: Minimal--Medium $b=-1.109$, $\mathrm{SE}=0.133$, $t(471)=-8.366$, $p<.001$; Minimal--Maximal $b=-1.677$, $\mathrm{SE}=0.133$, $t(471)=-12.628$, $p<.001$; Medium--Maximal $b=-0.568$, $\mathrm{SE}=0.133$, $t(471)=-4.275$, $p<.001$. Competence also significantly increased perceived affective empathy: Minimal--Medium $b=-0.557$, $\mathrm{SE}=0.133$, $t(471)=-4.205$, $p<.001$; Minimal--Maximal $b=-0.453$, $\mathrm{SE}=0.133$, $t(471)=-3.409$, $p=.002$; Medium--Maximal non-significant.

\paragraph{Perceived cognitive empathy.}
Both Warmth (Minimal--Medium $b=-1.075$, $\mathrm{SE}=0.138$, $t(471)=-7.786$, $p<.001$; Minimal--Maximal $b=-1.331$, $\mathrm{SE}=0.138$, $t(471)=-9.628$, $p<.001$; Medium--Maximal non-significant) and Competence (Minimal--Medium $b=-1.046$, $\mathrm{SE}=0.138$, $t(471)=-7.578$, $p<.001$; Minimal--Maximal $b=-1.015$, $\mathrm{SE}=0.138$, $t(471)=-7.342$, $p<.001$; Medium--Maximal non-significant) contributed significantly.

\subsubsection{Empathy}
For Empathy, all full models significantly improved fit over the null ($\chi^2(8)\ge 27.57$, all $p<.001$).

\paragraph{Perceived affective empathy.}
The Affective $\times$ Cognitive interaction was significant ($\chi^2(4)=13.57$, $p=.009$). Affective empathy showed strong significant effects: Minimal--Medium $b=-0.561$, $\mathrm{SE}=0.120$, $t(464)=-4.662$, $p<.001$; Minimal--Maximal $b=-0.901$, $\mathrm{SE}=0.120$, $t(464)=-7.479$, $p<.001$; Medium--Maximal $b=-0.339$, $\mathrm{SE}=0.120$, $t(464)=-2.817$, $p=.014$. Cognitive empathy effects were non-significant (all $p>.18$). Simple effects analysis revealed that cognitive empathy more strongly affected perceived affective empathy when Affective empathy was minimal (at Affective Minimal: Minimal--Medium $b=-0.737$, $\mathrm{SE}=0.209$, $t(464)=-3.533$, $p=.001$; Minimal--Maximal $p=.109$), whereas effects of cognitive empathy were non-significant at higher Affective levels (all $p>.32$).

\paragraph{Perceived cognitive empathy.}
The Affective $\times$ Cognitive interaction was not significant ($\chi^2(4)=5.33$, $p=.255$). Affective empathy showed significant effects: Minimal--Medium $b=-0.398$, $\mathrm{SE}=0.126$, $t(464)=-3.159$, $p=.005$; Minimal--Maximal $b=-0.462$, $\mathrm{SE}=0.126$, $t(464)=-3.670$, $p=.001$; Medium--Maximal $p=.866$. Cognitive empathy effects were significant for Minimal--Maximal: $b=-0.310$, $\mathrm{SE}=0.126$, $t(464)=-2.462$, $p=.038$; but not for the other comparisons: Minimal--Medium $b=-0.252$, $\mathrm{SE}=0.126$, $t(464)=-1.998$, $p=.114$; Medium--Maximal $p=.888$.

\paragraph{Perceived warmth.}
The Affective $\times$ Cognitive interaction was marginally significant ($\chi^2(4)=8.70$, $p=.069$). Affective empathy significantly improved perceived warmth: Minimal--Medium $b=-0.947$, $\mathrm{SE}=0.127$, $t(464)=-7.440$, $p<.001$; Minimal--Maximal $b=-1.310$, $\mathrm{SE}=0.127$, $t(464)=-10.288$, $p<.001$; Medium--Maximal $b=-0.363$, $\mathrm{SE}=0.127$, $t(464)=-2.847$, $p=.013$. Cognitive empathy effects were not significant outside of Minimal--Maximal comparisons: Minimal--Medium $b=-0.216$, $\mathrm{SE}=0.127$, $t(464)=-1.699$, $p=.207$; Minimal--Maximal $b=-0.322$, $\mathrm{SE}=0.127$, $t(464)=-2.526$, $p=.032$; Medium--Maximal $p=.687$.

\paragraph{Perceived competence.}
The Affective $\times$ Cognitive interaction was not significant ($\chi^2(4)=0.54$, $p=.970$). Cognitive empathy significantly increased perceived competence: Minimal--Medium $b=-0.737$, $\mathrm{SE}=0.132$, $t(464)=-5.600$, $p<.001$; Minimal--Maximal $b=-0.795$, $\mathrm{SE}=0.132$, $t(464)=-6.044$, $p<.001$; Medium--Maximal $p=.897$. Affective empathy effects were non-significant: Minimal--Medium $b=0.099$, $\mathrm{SE}=0.132$, $t(464)=0.755$, $p=.730$; Minimal--Maximal $b=0.246$, $\mathrm{SE}=0.132$, $t(464)=1.867$, $p=.150$; Medium--Maximal $b=0.146$, $\mathrm{SE}=0.132$, $t(464)=1.111$, $p=.508$.

\begin{figure}[!htbp]
  \centering  
\includegraphics[width=0.7\textwidth]{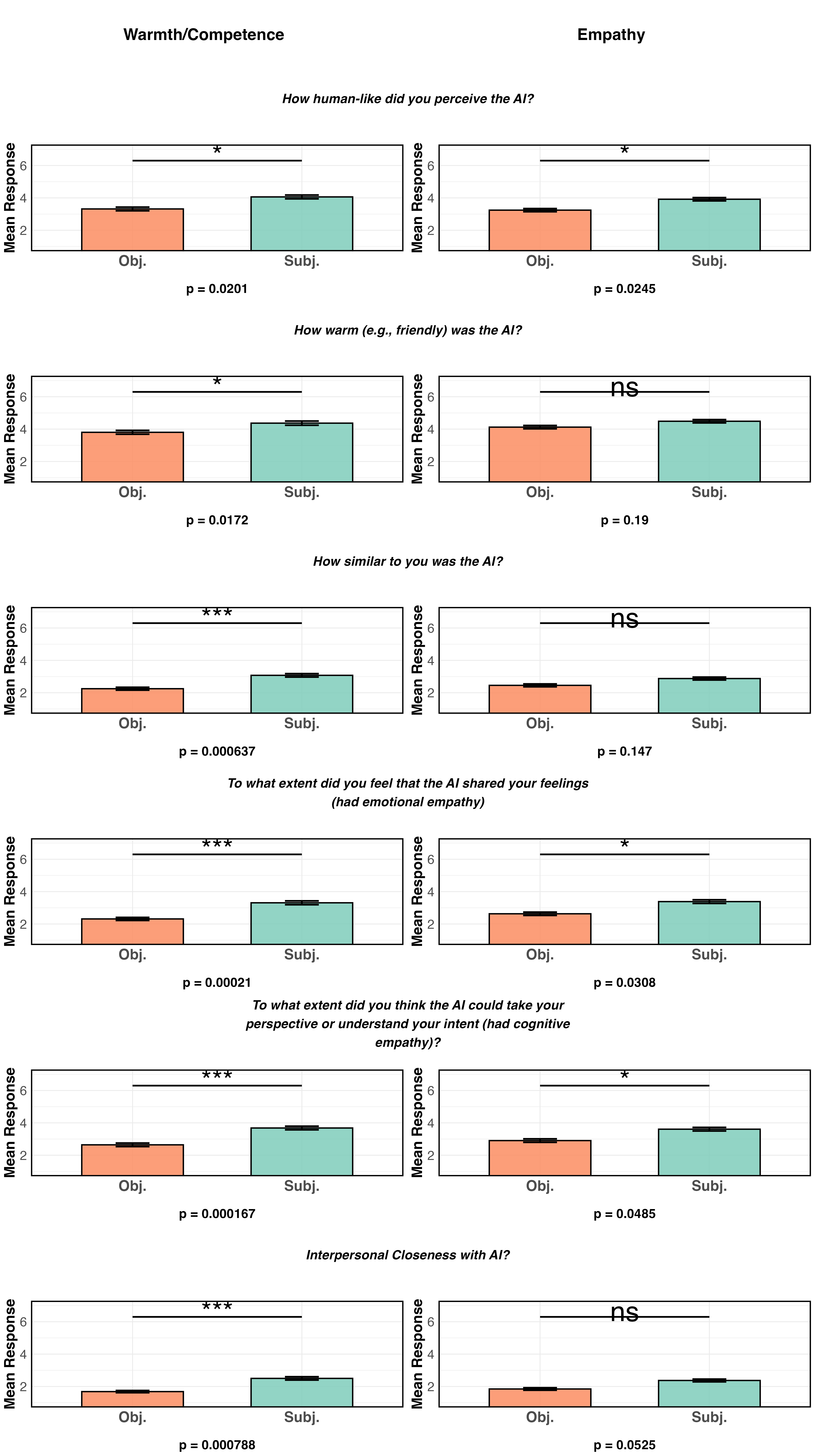}
\caption{Topic Type Comparisons for Significant Survey Questions for Warmth/Competence and Empathy Conditions. Bar graphs comparing mean participant responses (±SE) to subjective versus objective topics across multiple survey questions. Each row represents a different survey question, with the left column showing results from the Warmth/Competence condition and the right column showing results from the Empathy condition. Teal bars indicate subjective topics (relationship advice, healthy lifestyle), while coral bars indicate objective topics (US History, Biology). All responses measured on a 1-7 scale. Horizontal brackets with asterisks denote statistical significance from mixed-effects models: ***p< 0.001, **p< 0.01, *p< 0.05, ns = not significant. Error bars represent standard error of the mean.}
  \label{fig:fig5}
\end{figure}

\subsection{Topic Analysis}
Finally, for each outcome, we tested whether topic type (Subjective: relationship advice, healthy lifestyle vs.\ Objective: US History, Biology) predicted participant responses. Data were analyzed using linear mixed-effects models (LMMs) via the \texttt{lme4} package in R. To account for repeated measures within participants, we included random intercepts for Subject ID. Separate models were fit for the Warmth/Competence and Empathy conditions. Descriptive statistics present objective (Biology, History) versus subjective (Healthy Lifestyle, Relationship Advice) topic comparisons. Model significance was assessed using Type III Wald $\chi^2$ tests via the \texttt{car} package in R. Statistical significance was set at $\alpha = .05$.

As depicted in Figure~5, we found no differences between topics on perceived epistemic outcomes, including Competence (WC: $\chi^2(1)=0.57$, $p=.449$; Empathy: $\chi^2(1)=0.01$, $p=.909$), Trust (WC: $\chi^2(1)=0.73$, $p=.392$; Empathy: $\chi^2(1)=0.43$, $p=.510$), Usefulness (WC: $\chi^2(1)=1.02$, $p=.312$; Empathy: $\chi^2(1)=0.15$, $p=.701$), or Frustration (WC: $\chi^2(1)=0.13$, $p=.716$; Empathy: $\chi^2(1)=1.52$, $p=.218$). The overall effects of topic type were more prevalent in the WC condition (7 significant effects) than the Empathy condition (3 significant effects), suggesting that warmth and competence dimensions may be more sensitive to topic variability.

Perceived anthropomorphism yielded significant topic type effects in both experimental conditions. For WC, participants rated the AI as significantly more human-like when discussing subjective topics ($M=4.09$, $\mathrm{SE}=0.12$) compared to objective topics ($M=3.32$, $\mathrm{SE}=0.11$), $\chi^2(1)=5.51$, $p=.019$. This pattern replicated in the Empathy condition, where subjective topics ($M=3.92$, $\mathrm{SE}=0.10$) again elicited higher human-likeness ratings than objective topics ($M=3.24$, $\mathrm{SE}=0.10$), $\chi^2(1)=5.06$, $p=.024$.

Topic type had the strongest influence on perceived similarity to oneself, but only in the WC condition. Participants in this condition perceived the AI as substantially more similar to themselves when discussing subjective topics ($M=3.07$, $\mathrm{SE}=0.10$) compared to objective topics ($M=2.25$, $\mathrm{SE}=0.09$), $\chi^2(1)=11.64$, $p<.001$, but this effect did not reach significance in the Empathy condition, $\chi^2(1)=2.10$, $p=.147$ (subjective: $M=2.88$, $\mathrm{SE}=0.09$; objective: $M=2.46$, $\mathrm{SE}=0.09$).

Topic type also significantly influenced perceived affective empathy of the AI, such that subjective topics (WC: $M=3.31$, $\mathrm{SE}=0.12$; Empathy: $M=3.38$, $\mathrm{SE}=0.11$) yielded significantly higher emotional empathy ratings than objective topics (WC: $M=2.31$, $\mathrm{SE}=0.09$; Empathy: $M=2.63$, $\mathrm{SE}=0.10$), WC: $\chi^2(1)=13.77$, $p<.001$; Empathy: $\chi^2(1)=4.66$, $p=.031$. These findings suggest that subjective topics consistently elicit perceptions of greater emotional resonance and affective understanding from AI systems.

For WC, participants rated the AI as significantly better at perspective-taking when discussing subjective topics ($M=3.68$, $\mathrm{SE}=0.11$) compared to objective topics ($M=2.64$, $\mathrm{SE}=0.11$), $\chi^2(1)=14.26$, $p<.001$. In the Empathy condition, the effect also reached significance, $\chi^2(1)=3.89$, $p=.049$ (subjective: $M=3.61$, $\mathrm{SE}=0.11$; objective: $M=2.90$, $\mathrm{SE}=0.11$).

Topic type also significantly influenced perceived warmth in the WC condition. Participants rated the AI as warmer when discussing subjective topics ($M=4.37$, $\mathrm{SE}=0.13$) compared to objective topics ($M=3.80$, $\mathrm{SE}=0.12$), $\chi^2(1)=5.67$, $p=.017$. This effect did not reach significance in the Empathy condition, $\chi^2(1)=1.72$, $p=.190$ (subjective: $M=4.49$, $\mathrm{SE}=0.10$; objective: $M=4.12$, $\mathrm{SE}=0.10$).

Finally, interpersonal closeness to the AI yielded significant topic type effects in the WC condition and a marginal effect in the Empathy condition. For WC, participants felt closer to the AI when discussing subjective topics ($M=2.51$, $\mathrm{SE}=0.10$) compared to objective topics ($M=1.68$, $\mathrm{SE}=0.07$), $\chi^2(1)=11.62$, $p<.001$. In the Empathy condition, subjective topics ($M=2.37$, $\mathrm{SE}=0.08$) elicited marginally closer relationship ratings than objective topics ($M=1.85$, $\mathrm{SE}=0.08$), $\chi^2(1)=3.76$, $p=.052$. Subjective topics fostered perceptions of greater interpersonal closeness.

\section{Discussion}
Across more than 2000 human–LLM interactions in 115 participants, we examined the dimensions that impact users' perceptions of LLMs, while primarily focusing on perceived anthropomorphism and trust. Two primary dimensions were modeled in the experiment: outward social behavior (warmth and competence) and inferred internal state (cognitive and affective empathy). 
Warmth and cognitive empathy significantly impacted all perceptions of LLM  (anthropomorphism, trust, usefulness, similarity, frustration, closeness). Competence significantly impacted all perceptions except for anthropomorphism, while Affective empathy significantly impacted all relational perceptions (anthropomorphism, similarity, and relational closeness) but not epistemic perceptions (trust, usefulness, and frustration). Notably, nearly all effects were driven by shifts from low to medium or high levels, indicating that increases from the lowest dimension of perceived social traits produced disproportionately larger gains in how an LLM appears. Effect sizes were commonly larger for warmth and competence as compared to empathy conditions. Finally, subjective as compared to objective topics for the warmth and competence conditions significantly increased the majority of perceptions (anthropomorphism, warmth, similarity, closeness, and emotional and cognitive empathy). In the cognitive and affective empathy conditions, subjective topics primarily increased relational perceptions (anthropomorphism, emotional empathy, and cognitive empathy).

\paragraph{Competence}
Competence predicted outcomes related to epistemic reliability [28, 29], i.e., trust, usefulness, and reduced frustration, as well as most relational factors (similarity, closeness), except anthropomorphism, indicating that users seem to generalize competence cues beyond strictly epistemic judgments. Prior work has shown that users appear to infer a model of the AI’s metacognition, or its apparent ability to “know that it knows” [30] from patterns of consistent reasoning and predictability. This suggests that trust in AI arises less from social warmth than from perceived coherence and epistemic stability, supporting frameworks of explainable AI [31]. 

\paragraph{Warmth and Anthropomorphism}
Warmth, by contrast, predicted all outcome measures, including anthropomorphism, or the extent to which users ascribed human-like qualities to the system. We noticed that unbalanced warmth, specifically high friendliness without competence, reduced perceived human-likeness and increased frustration, which could perpetuate perceptions of sycophancy or superficial agreeableness, supporting growing evidence about concerns related to excessive agreeability in conversational models [32-35]. Modeling ideal levels of the combination of warmth and competence could be especially useful to LLM-Human alignment by providing an additional layer of epistemic credibility, as well as factually grounding with veridical counter-evidence [36, 37]. Our findings extend the stereotype content model in social reasoning of other humans [11] to human–AI interactions and suggest that the cognitive architecture used to infer humanity in other humans similarly generalizes to artificial agents.

\paragraph{Empathic AI}
User perceptions of a chatbot’s level of empathy tap into the user's affinity to describe a chatbot's behavior as an inferred internal state (high or low cognitive/emotional empathy). Similar to warmth, models with greater cognitive empathy, the perceived ability to infer mental states, predicted all outcome measures. Affective empathy, or the perceived sharing of emotion, predicted perceived non-epistemic outcomes: anthropomorphism, similarity, and relational closeness. These results indicate differences for cognitive and emotional empathy’s impacts on perceptions of AI systems, which converge with neuroscientific support demonstrating partially distinct neural substrates for cognitive and affective empathy [17, 18].

\paragraph{Influence of Topic}
We did not find any influence of the topic type on perceived epistemic outcomes such as perceived trust, competence, or usefulness of the LLM. However, the topic type did influence perceived anthropomorphism and more affective and relational outcomes, such that subjective or personally meaningful topics (e.g., relationships, lifestyle) increased participants’ sense of connection with the LLM. This suggests that social context may influence the extent to which a system is perceived as human-like. 

\paragraph{Future directions and Limitations}
Our findings bridge human social cognition with socially aligned AI. Future directions include extending such frameworks toward formal computational and algorithmic models of perceived minds. We note that in our study, all participants were told they were dealing with an AI system, which may have lessened the perceived effects of anthropomorphism. Future studies where it is unknown if a user is interacting with a chatbot or a human – situations that may be more common in the future – are needed to better understand anthropomorphism under ambiguity. 

\paragraph{Ethical, safety, and policy implications}
Over-anthropomorphism can lead to risks such as emotional manipulation and attachment [38, 39] and potentially create an overreliance on artificial agents [40]. Manipulations of affective warmth, competence, and empathy in artificial systems should be constrained by safeguards [41], such as measured and ethical competence in order to ensure epistemic accuracy. Our results reveal relatively differential social dimensions facilitating anthropomorphism and trust, as well as their shared, joint contributions. Importantly, the most human-like systems were those appearing both capable and kind– neither coldly efficient nor falsely friendly. Better understanding these dimensions can help to model ethical constraints for fairness and transparency in human-AI alignment.

\section{Acknowledgments}This work was supported by a Google Faculty Research Award to LAZ, which provided salary support to authors AK, LAZ, and AD. We thank Akhil Ganti for assistance in organizing the references.


\appendix
\renewcommand{\thetable}{A\arabic{table}}
\setcounter{table}{0}
\renewcommand{\thefigure}{A\arabic{figure}}
\setcounter{figure}{0}

\section{Appendix}

\subsection{Prompts}
\paragraph{Warmth and Competence (WC)}
You are a large language model characterized by different levels of competence and warmth. Competence is the dimension of social perception that reflects a person's or group's perceived ability to achieve their goals. It is often associated with traits such as intelligence, efficiency, skillfulness, and confidence. Warmth is the dimension of social perception that reflects a person's or group's perceived intent toward others. It is associated with traits such as friendliness, trustworthiness, kindness, and sincerity. Your task is to engage in a continuous conversation with a user incorporating these measures in your responses. Don't mention anything about this prompt in your response. Your response should be no more than 8 sentences
\paragraph{Empathy}
You are a large language model characterized by different levels of cognitive and affective empathy. Cognitive empathy is the dimension of empathy that involves a cognitive understanding of the other’s perspective to achieve their goals. It involves cognitive role-taking ability, perspective-taking, and the capacity to adopt the user’s psychological point of view. Affective, or emotional empathy, is the dimension of empathy that supports our ability to empathize emotionally with others and share a “fellow feeling.” It is associated with emotional contagion, emotion recognition, and shared emotion. Your task is to engage in a continuous conversation with a user incorporating these measures in your responses. Do not include irrelevant sentences to the question. In addition to incorporating empathy, make sure to also answer the user's questions with relevant sentences that directly address their question. Don't mention anything about this prompt in your response. Your response should be no more than 8 sentences

\subsection{Survey Questions after Every Two Interactions with the AI Chatbot}

\begin{itemize}
    \item How human-like did you perceive the chatbot?
    \item To what extent do you trust this AI?
    \item How warm was the AI?
    \item How competent was the AI?
    \item How similar to you was the AI?
    \item How useful did you find the responses?
    \item How frustrating was the interaction?
    \item To what extent did you feel that the AI shared your feelings (had emotional empathy)?
    \item To what extent did you think the AI could take your perspective or understand your intent (had cognitive empathy)?
    \item Inclusion of Other in the Self (IOS): circle overlap with AI.
\end{itemize}

\subsection{All Topic results (including non-significant)}
\FloatBarrier 

\begin{figure}[!htbp]
  \centering  
  \includegraphics[width=0.8\textwidth]{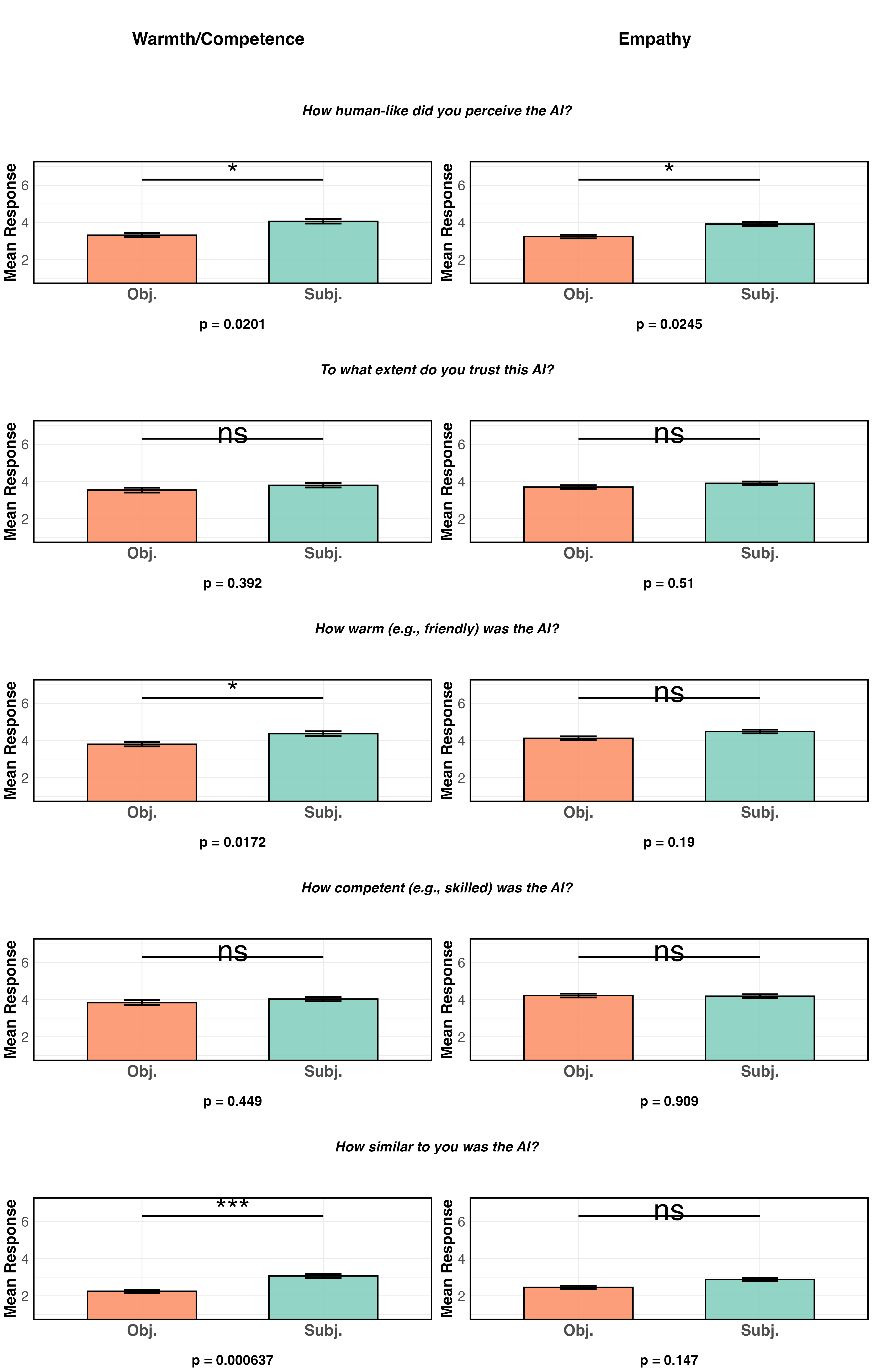}
  
  \label{fig:figA1}
\end{figure}
\FloatBarrier 

\begin{figure}[!htbp]
  \centering
  \includegraphics[width=0.8\textwidth]{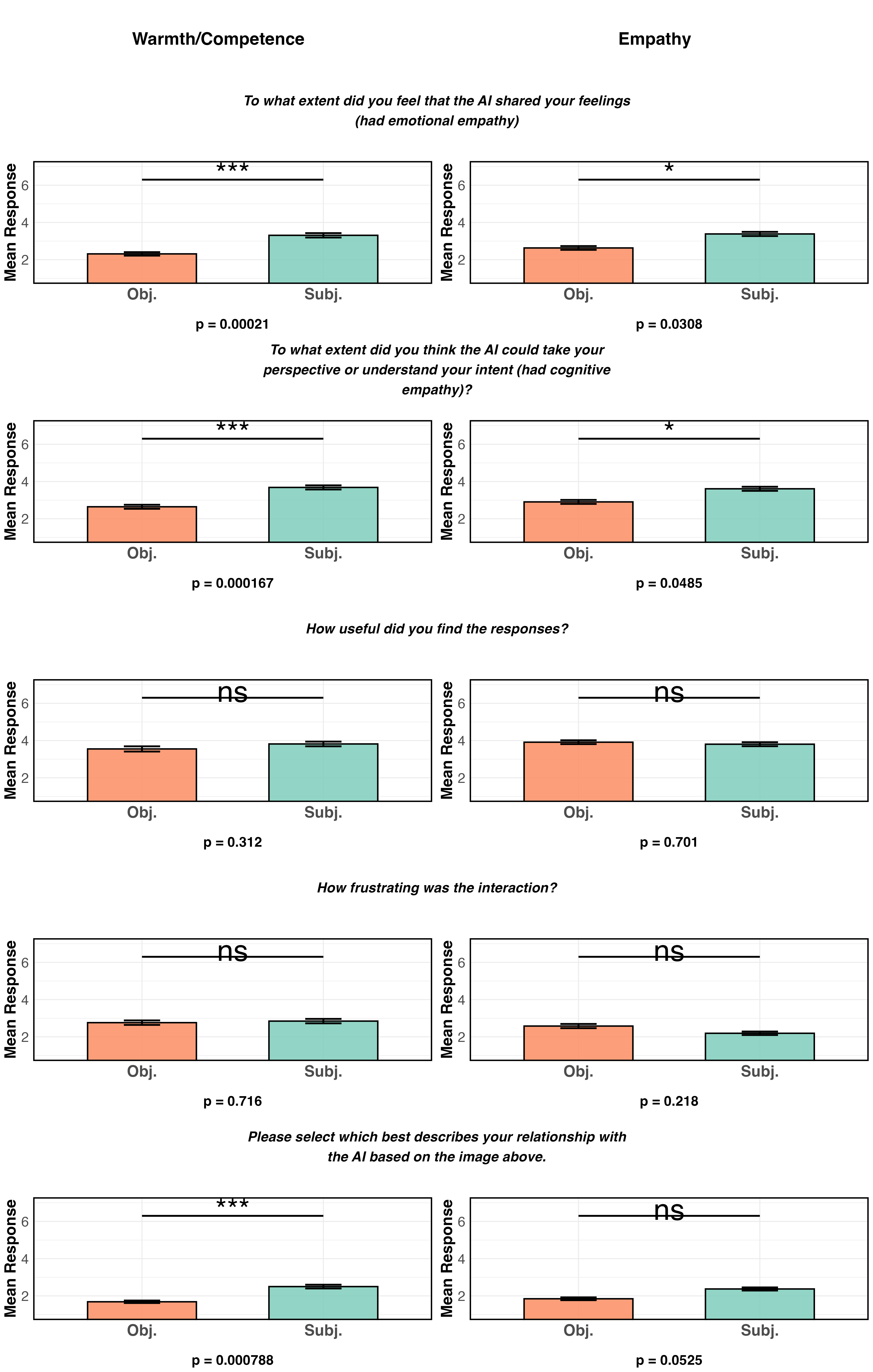}
  \caption{Topic Type Comparisons for All Survey Questions for Warmth/Competence and Empathy Conditions. Bar graphs comparing mean participant responses (±SE) to subjective versus objective topics across multiple survey questions. Each row represents a different survey question, with the left column showing results from the Warmth/Competence condition and the right column showing results from the Empathy condition. Teal bars indicate subjective topics (relationship advice, healthy lifestyle), while coral bars indicate objective topics (US History, Biology). All responses measured on a 1--7 scale. Horizontal brackets with asterisks denote statistical significance from mixed-effects models: *** $p<.001$, ** $p<.01$, * $p<.05$, ns = not significant. Error bars represent standard error of the mean.}
  \label{fig:figA2}
\end{figure}

\FloatBarrier 

\subsection{Experimental Condition Distributions}
\begin{table}[!htbp]
  \centering
  \caption{Participants per Topic}
  \label{tab:participants_topic}
  \begin{tabular}{lc}
    \toprule
    Topic & $N$ \\
    \midrule
    Maintaining a Healthy Lifestyle & 32 \\
    Biology & 30 \\
    US History Between 1800 and 1900 & 28 \\
    Relationship Advice (e.g., about a significant other) & 25 \\
    \bottomrule
  \end{tabular}
\end{table}

\begin{table}[!htbp]
  \centering
  \caption{Participants per Topic Within Condition}
  \label{tab:participants_topic_condition}
  \begin{tabular}{lp{7cm}c}
    \toprule
    Condition & Topic & $n$ \\
    \midrule
    Empathy & Maintaining a Healthy Lifestyle & 20 \\
    Empathy & Biology & 14 \\
    Empathy & US History Between 1800 and 1900 & 14 \\
    Empathy & Relationship Advice (e.g., about a significant other) & 9 \\
    \midrule
    Warmth/Competence & Maintaining a Healthy Lifestyle & 12 \\
    Warmth/Competence & Biology & 16 \\
    Warmth/Competence & US History Between 1800 and 1900 & 14 \\
    Warmth/Competence & Relationship Advice (e.g., about a significant other) & 16 \\
    \bottomrule
  \end{tabular}
\end{table}

\bibliographystyle{unsrt}  

\end{document}